\documentclass[rnote]{aa}  
\usepackage{graphicx}
\usepackage{txfonts}
\usepackage{natbib}
\bibpunct{(}{)}{;}{a}{}{,}

\begin{document}
   \title{Long-Term Optical Monitoring of $\eta$~Carin\ae}

   \subtitle{Multiband light curves for a complete orbital period}

   \authorrunning{Fern\'andez-Laj\'us et al.}

   \author
	{
	E.~Fern\'andez-Laj\'us    \inst{1,2,3}
	 \fnmsep\thanks{\email{eflajus@fcaglp.unlp.edu.ar}},
	C.~Fari\~na     \inst{1,2,3},
        A.F.~Torres     \inst{1,2,3},
        M.A.~Schwartz     \inst{1},
        N.~Salerno      \inst{1},
        J.P.~Calder\'on \inst{1},\\
        C.~von Essen    \inst{1},
	L.M.~Calcaferro   \inst{1},
        F.~Giudici      \inst{1},
        C.~Llinares     \inst{4}
        \and
        V.~Niemela       \inst{5}
         }

   \offprints{Eduardo Fern\'andez-Laj\'us}

   \institute{
              Facultad de Ciencias Astron\'omicas y Geof\'{\i}sicas (FCAG) -
              Universidad Nacional de La Plata (UNLP)\\
              Observatorio Astron\'omico, Paseo del Bosque S/N - 1900 La Plata, Argentina
	\and
	      Instituto de Astrof\'{\i}sica de La Plata (CONICET-UNLP), Argentina
         \and
             Fellow of CONICET, Argentina
         \and
             Astrophysical Institute Potsdam, An der Sternwarte 16, Germany
         \and
	      In Memoriam (1936-2006)
             }

   \date{Received July 29, 2008; accepted September 16, 2008}

 
  \abstract
   {
The periodicity of 5.5 years for some observational events occurring in
$\eta$~Carin\ae\/ manifests itself across a large wavelength range
and has been associated with its binary nature. 
These events are supposed to occur
when the binary components are close to periastron. 
To detect the previous periastron passage of $\eta$~Car in 2003, we started an
intensive, ground-based, optical, photometric observing campaign.
   }
   {	
We continued observing the object to monitor its
photometric behavior and variability across the entire orbital cycle. 
   }
   {
Our observation program consisted of daily differential photometry from CCD 
images, which were acquired using a 0.8 m telescope and a standard $BVRI$
filter set at La Plata Observatory.
The photometry includes the central object and the surrounding
    Homunculus nebula.
   }
   {	
We present up-to-date results of our observing program, 
including homogeneous photometric data collected between 2003 and 2008. 
Our observations demonstrated that $\eta$~Car has continued increasing in brightness at
a constant rate since 1998. 
In 2006, it reached its brightest magnitude ($V \sim 4.7$) since about 1860s.
The object then suddenly reverted its brightening trend, fading to $V = 5.0$ at the
beginning of 2007, and has maintained a quite steady state since then.
We continue the photometric monitoring of $\eta$~Car in anticipation of
the next ``periastron passage'', predicted to occur at the beginning of 2009.
}


   \maketitle
%
\section{Introduction}

$\eta$~Carin\ae\ is the brightest ($V\sim 5$) Luminous Blue Variable 
star in the sky.
The central object is surrounded by a nebulosity, a product of the
mass ejection that occurred during the {\em``Great Eruption''} in the first
half of the 19th century.
It consists of a $18''$ long axis bipolar reflection nebula, which was called ``Homunculus'' by \citet{Gaviola}.

The central star is suspected to be a binary system as first
proposed by \citet{D97}, from the evidence of the 
5.5 years periodicity of the ``spectroscopic events''.
These events consist of the attenuation or complete disappearance of high
excitation lines \citep{D96}.

The most recent ``spectroscopic event'' occurred in 2003.5 and, for that
  reason, $\eta$~Car was the target of an intensive multiwavelength
  observational campaign.
Clear dips were registered in X-ray luminosity \citep{C05},   
near-infrared $JHKL$ photometry \citep{White04}, and
7-mm \citep{Abraham05a} and 3-cm wave emissions \citep{Duncan03}. 
In the optical range, we carried out a multicolor CCD monitoring of
$\eta$~Car.
Our observations started in January 2003, and we detected a minimum in the four $BVRI$ bands \citep[hereafter Paper I]{ibvs5477}, more than a week after detecting a minimun in X-ray luminosity. 
These sudden brightness fadings are occasionally referred to as ``eclipse-like'' events \citep[e.g.][]{White04}, and are explained well 
by a close binary scenario, in which the hot secondary star moves in a highly 
eccentric orbit of period 2022.7 days \citep{Damineli08a}.
Many of these events are supposed to be produced during periastron passages
\citep{D97} and extend over only a few months due to the
high orbital eccentricity. Nevertheless, some other features vary continuously throughout the complete cycle \citep[e.g.][]{Duncan03,Damineli08b}   . 

Following the 2003 observing campaign, we continued monitoring the
photometric behavior during the complete orbital period, including the upcoming
``eclipse-like'' event, predicted for next January 2009.
It is worth emphasizing the significant advantage of the daily availability of a telescope of the appropriate size, and retaining the same instrumental 
configuration (i.e. ``telescope + filter-set + detector'').
This is a desirable factor when high-precision is sought in long-term ground-based photometric monitoring of $\eta$~Car  \citep[and references therein]{Sterken01}. 

 In this paper, we present the results of our long-term $BVRI$ CCD differential
 photometry of $\eta$~Car, performed between 2003 and 2008, 
covering the entire orbital period.

\section{Observations}
\subsection{Images acquisition}
 Our observing program consists of the daily acquisition of digital images (weather
permitting) during the annual observational seasons. Each season starts in the
middle of November of the preceding year and finishes at the end of August
(for details see Table~\ref{seasons}).  
Observing seasons are separated by gaps of about 75 days, when the $\eta$~Car
position in the sky was beyond the telescope pointing limits or the airmass
exceeded reasonable values.
\begin{table}
\caption{Dates and Julian Day Numbers of the beginning and end of our annual
  $\eta$ Car observing seasons.}
\label{seasons}
\centering    
\begin{tabular}{c | c c | c c}
\hline\hline               
Observing&\multicolumn{2}{c | }{Start}&\multicolumn{2}{c}{End}\\
Season 	& Date &  JDN & Date &  JDN  \\
\hline                        
$2003^{\dag}$ & Jan 17, 2003 & 2452656 & Aug 29, 2003 & 2452881 \\
2004 & Nov 14, 2003 & 2452957 & Aug 27, 2004 & 2453244 \\
2005 & Nov 17, 2004 & 2453326 & Aug 30, 2005 & 2453612 \\
2006 & Nov 16, 2005 & 2453690 & Aug 28, 2006 & 2453976 \\
2007 & Nov 13, 2006 & 2454052 & Aug 31, 2007 & 2454344 \\
2008 & Nov 23, 2007 & 2454428 & Aug 31, 2008 & 2454710 \\

\hline        
\noalign{\smallskip}
\multicolumn{5}{l}{\small (\dag) Photometric data already published in PAPER I}
\end{tabular}
\end{table}
Image acquisition was performed using a CCD camera mounted on the 0.8 m
``Virpi S. Niemela'' (VSN) telescope (f/20.06 Cassegrain), with a
Johnson-Cousins $BVRI$ filter set, at La Plata Observatory\footnote{La Plata
  Observatory belongs to FCAG-UNLP}, Argentina.  
The $BVRI$ passbands used were those recommended by \citet{Bessell} for coated CCDs. 
The camera is a Photometrics STAR I containing a Thomson TH7883PS coated
scientific-grade, front-illuminated CCD chip. 
The chip array is 384 x 576 pixels (23 $\mu$m square pixel), giving
1$.\!{\!^{'}}$9 x 2$.\!{\!^{'}}$8 field images at the telescope focal plane,
the scale being 0$.\!{\!^{''}}$296 per pixel. 
In Fig. 1, a typical frame is shown, in which the main stars employed for the
photometry are identified. 
HDE303308 ($V=8.15$), located at about $1'$ NNE of $\eta$~Car, is used as
comparison star. 
CPD-59~2627 (= Tr16-3\footnote{Following the nomenclature of \citet{FMM73}}, $V=10.11$) and CPD-59~2628 (= Tr16-1, $V=9.56$)
are considered to check the $\eta$~Car photometry. 
The nearby stars Tr16-64 ($V = 10.72$), Tr16-65 ($V = 11.09$), and Tr16-66
($V=11.98$), located about $15''$-$20''$ from the center of $\eta$~Car are
also labeled in the figure.

\begin{figure}[!t]
\center
{\includegraphics[height=12.5cm,angle=0]{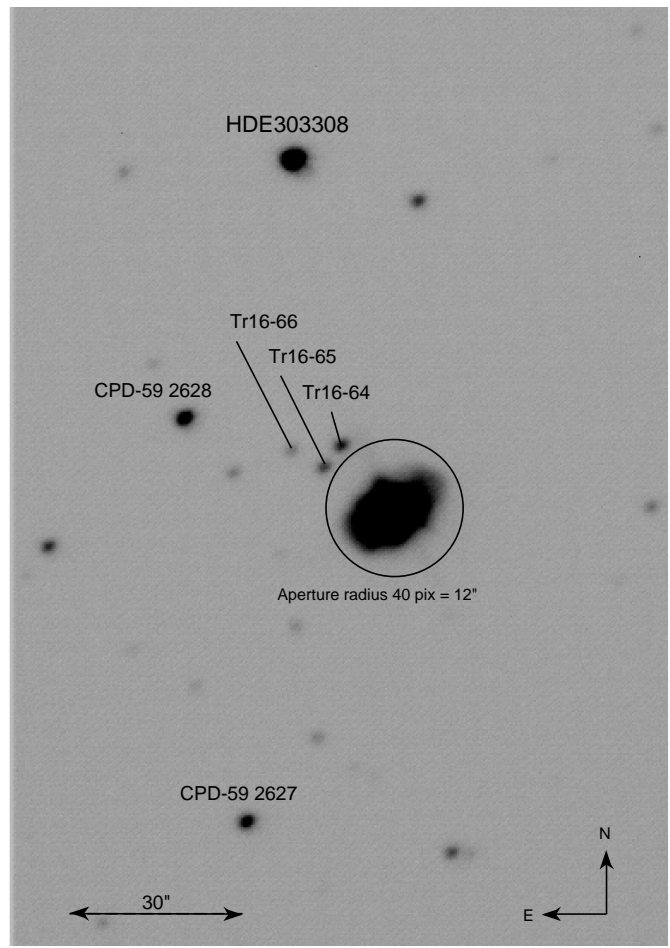}}
\caption{
One of our $V$ images of the $\eta$~Car frame. 
The location of the comparison star HDE303308 and checking stars 
CPD-59 2628 and CPD-59 2627 are indicated, as those faint but very near
stars Tr16-64, Tr16-65 and Tr16-66. 
The figure also depicts the 40 pix = $12''$ radius aperture
considered to extract the instrumental magnitudes of $\eta$~Car.
}
\label{frame}
\end{figure}

Every night, a series of images were acquired during a time interval of no 
longer than 30 minutes, for each filter. Each series comprised typically of about 10 or more images. 
In this way, more than 23\,000 images of $\eta$~Car were obtained using $B, V,
R,$ and $I$ filters during the 2004 to 2008 observing seasons,
consisting of more than 26\,000 images since the beginning of the campaign in
January 2003.
Whenever possible, bias, dark, and flat-field frames were acquired.
However, since calibration images were not available for all nights of the
campaign, in this work we used only uncalibrated science frames.

\subsection{Data reduction}
Instrumental magnitudes of each star were determined by means of aperture
photometry. 
After Paper I, a new image-processing pipeline was written to
complete aperture photometry for the brightest stars included in the
$\eta$~Car frame.
This pipeline was developed by using the {\small IRAF}
\footnote{IRAF is distributed by the NOAO, operated by the AURA, Inc, under
cooperative agreement with the NSF, USA.}
command language and other tasks such as the {\small APPHOT} photometry package.
With this tool, we proceeded to measure the images acquired during the entire
campaign, including those of the 2003 observing season.

We fixed the aperture radius at 40 pix = $12''$ to extract the instrumental magnitude of $\eta$~Car and ensure that we measured both the integrated
flux of the central object and the major part of the Homunculus nebula.
 This aperture also avoids the contribution of light from the
neighboring stars Tr16-64, Tr16-65, and Tr16-66 (see Fig. 1).
After the instrumental magnitudes had been obtained, we calculated differential
magnitudes relative to HDE303308. This object was selected because it was the
nearest bright star to $\eta$~Car and constant in brightness \citep{Sterken01, Frey}. We used the same aperture for
$\eta$~Car as for HDE303308, to prevent spurious
variations caused by changes in the size or shape of the stellar intensity
profiles due to turbulence, bad focus, or tracking failures. 

We computed the mean differential magnitudes of $\eta$~Car to be the weighted average
of the relative magnitudes measured in each image belonging to the same
series. The weights used for these averages were derived from the errors in individual measurements for each image. 
A 2$\sigma$~($\sigma =$ std dev) rejection
criteria was then applied, and revised mean values were calculated. 
The light curves were constructed from these mean values, and their standard
deviations were adopted as a measure of the errors ($\epsilon$).
The averaged errors of our differential photometry were:
$\epsilon_B=\pm 0.010$, 
$\epsilon_V=\pm 0.007$, 
$\epsilon_R=\pm 0.012$, and
$\epsilon_I=\pm 0.015$ mag. 

To obtain a homogeneous data set for the entire campaign, we measured again all images taken during the 2003 observing season, using the aperture size mentioned
above. We compared these results with those published in Paper I, which were
extracted using a larger aperture radius i.e. 75 pix = $22''$. 
An excellent correlation between both data sets was verified. 
Nevertheless, in further analysis, we considered the small aperture to be the superior choice, because it eliminates any contamination from the neighboring stars, and
reduces the background noise that would be introduced by a larger aperture.
Furthermore, to estimate the seeing for each image, we measured the
FWHM of the stellar profile of HDE303308. A typical FWHM for the entire
campaign was about $3''$. 

The procedure for determining the mean differential magnitudes of the check
stars relative to HDE303308 was exactly the same as described above for
$\eta$~Car, apart from that a smaller aperture radius (10 pix = $3'' \sim 1~$FWHM)
was considered for extracting the instrumental magnitudes.
Since images of  both stars were always underexposed because the exposure times were
optimized for the light flux of $\eta$~Car, we chose this aperture to
maximize the signal-to-noise ratio. 

The check star CPD-59~2627 has exhibited an almost constant flux throughout the entire campaign,
in agreement with the results of \citet{Frey}.
The other check star, CPD-59~2628, is an eclipsing binary \citep[cf.][]{Frey}, for which we reproduced a light curve quite accurately. 
For this star, we also developed a numerical eclipsing-binary model using the
Wilson-Devinney Code \citep[e.g.][and references therein]{WvH04}. 
The difference (O-C) between our observational data and the model provides
us with a control tool for the $\eta$~Car photometry. The standard deviation
of these differences is about 0.02 mag in all bands.

\section{Results}
\begin{figure*}
\centering
        {\includegraphics[width=18.0cm,angle=0]{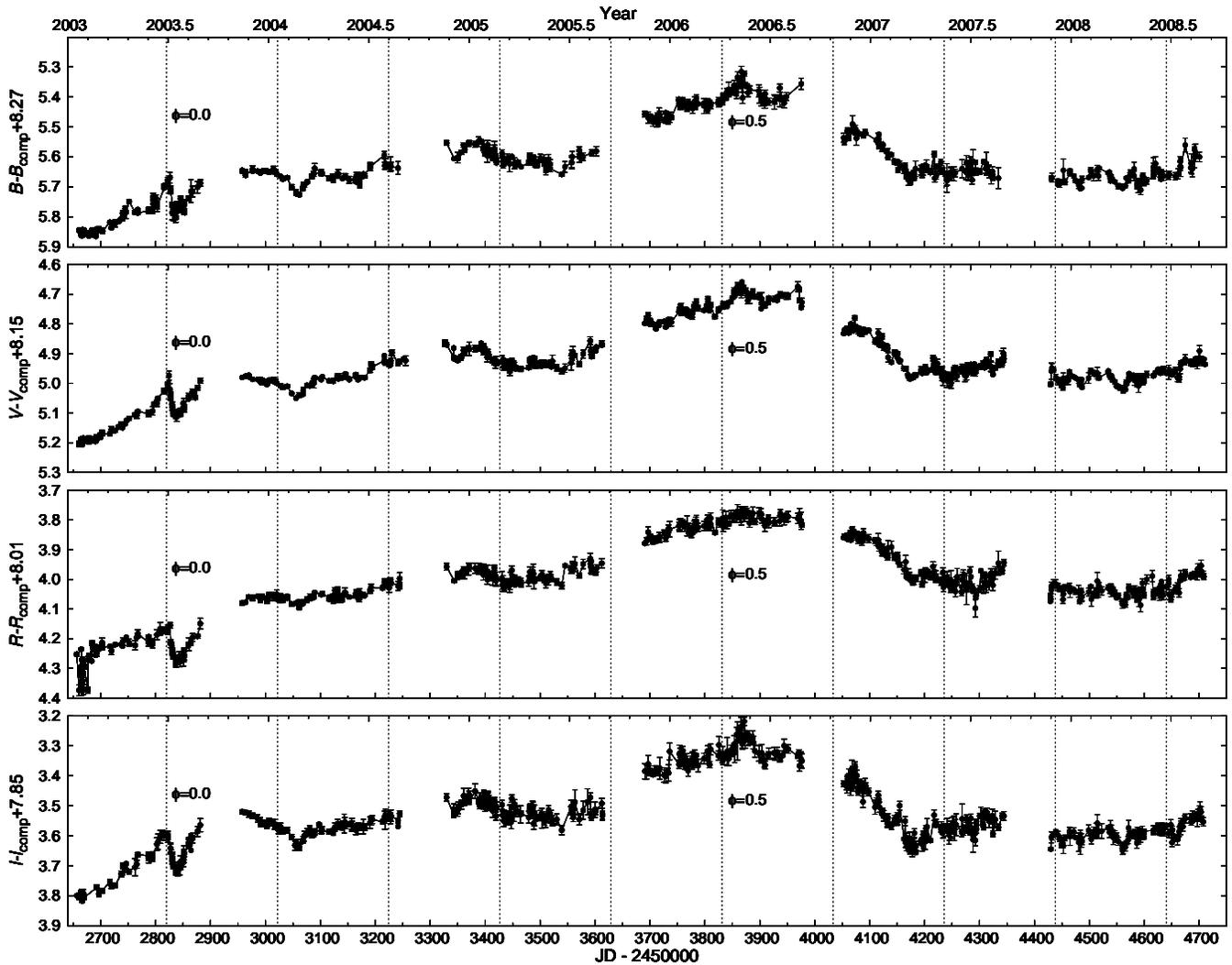}}
        \caption{
	$BVRI$ light curves of $\eta$~Car observed between January 17,
        2003 and August 31, 2008. The photometric values of HDE303308 taken
        from \citet{Feinstein82} are used as zeropoints.
        Bars represent the standard deviations of the mean values.
        The vertical grid indicates the orbital phases in steps of 0.1,
        according to the ephemeris 
        given by \citet{Damineli08a}. Phases $\phi = 0.0$ and $\phi = 0.5$
        are labeled.
        The different annual observing seasons are clearly delimited by gaps.
        }
        \label{plots}
\end{figure*}

In Fig.~\ref{plots}, we show the light curves
resulting from our $BVRI$ differential photometry of $\eta$~Car during the 2004-2008 observing seasons. 
For completeness, we also include the 2003 observing season data.
In this figure, we have used as zeropoints the $UBVRI$ Johnson-Kron-Cousins
photometry of HDE303308 provided by \citet{Feinstein82}, i.e. $B = 8.27, V = 8.15, 
R = 8.01$, and $I = 7.85$.
Top axis depicts the year and the vertical lines represent different
values of the orbital phase ($\phi$), separated by an increment of 0.1. 
The phase values are derived from the ephemeris ~$JD_{min} = 2452819.8 +
2022.7 E$ given by \citet{Damineli08a}, which corresponds to the time of
minimum of the narrow component of HeI $\lambda$6678 line.

Each light curve consists of more than 600 points. 
These data are available as an electronic table at the CDS 
\footnote{Centre de Donn\'ees astronomiques de Strasbourg, 
http://webviz.u-strasbg.fr/viz-bin/VizieR}.
An excerpt from it is presented in Table~\ref{data},
which contains, in successive columns: the JD, the band,
the mean differential magnitude, and its rms error of $\eta$ Car,
CPD-59~2627, and CPD-59~2628, the FWHM (in arc sec) determined
for HDE303308, and the number of images used to evaluate the mean values.
\begin{table*}
\caption{
Relevant data of our $BVRI$ differential photometry of $\eta$~Car against
HDE303308. 
This is an extract of the electronic table, where we list the Julian Date
(JD), the band (Bd), the mean differential magnitude
($\Delta$m), and the rms error ($\epsilon$) of $\eta$ Car, CPD-59~2627, and
CPD-59~2628, the full width half maximum of HDE303308
(FWHM), and the number of images considered in deriving the mean values (n).
} 
\label{data}     
\centering                       
\begin{tabular}{c | c | c c | c c | c c | c | c}      
\hline\hline                
   &  &\multicolumn{2}{c|}{$\eta$~Car}&\multicolumn{2}{c|}{CPD-59
     2627}&\multicolumn{2}{c|}{CPD-59 2628}&HDE303308&\\ 
JD & Bd &$\Delta$m & $\epsilon$ & $\Delta$m & $\epsilon$&
$\Delta$m & $\epsilon$ & FWHM ($''$)& n \\  
\hline                       
2454311.4378 & $B$ & -2.630 & 0.015 & 2.019 & 0.071 & 1.342 & 0.014 & 3.5 & 10 \\
2453601.4763 & $V$ & -3.259 & 0.004 & 1.965 & 0.007 & 1.334 & 0.020 & 3.2 & 10 \\
2452876.4586 & $R$ & -3.818 & 0.007 & 1.912 & 0.011 & 1.341 & 0.011 & 3.7 & 11 \\
2454130.6201 & $I$ & -4.307 & 0.009 & 1.897 & 0.018 & 1.366 & 0.014 & 2.9 &  9 \\
\hline                                 
\end{tabular}
\end{table*}

All bands exhibit similar behavior throughout the entire campaign.
The 2003.5 event is clearly appreciable as an increasing light phase which
starts at about JD 2452790 ($\phi \sim -0.015$), or even before.
After reaching maximum light at JD 2452825 ($\phi \sim 0.0026$), the brightness declines suddenly by about 0.13 mag, reaching a minimum 12 days later (JD 2452837, $\phi \sim 0.01$).
Following this photometric event, a broad maximum is spread over 218 days, 
reaching a second minimum at around JD 2453055 ($\phi \sim 0.12$).
This minimum appears to be present in the same orbital phase, two
and four cycles behind, if we consider the data published by van
\citet{vG06,Sterken99}.
Gradually, the object recovers the global brightening tendency exhibited since
2003.0, lasting 2006.35. 
The brightening rate computed was typically $0.13~\rm{mag~yr}^{-1}$
($0.12, 0.13, 0.14$, and $0.12~\rm{mag~yr}^{-1}$ for $B, V, R$, and $I$
respectively).
This secular brightening rate is almost identical to that $\eta$~Car displayed
during the 1997-1999 brightening \citep[cf.][]{Sterken99,Davidson99,MK04}.

After about JD 2453540,
$\eta$~Car's brightness increases to a maximum at about JD 2453860 
($\phi \sim 0.51$), the magnitude being $\sim 4.7$ in the $V$ band. The V-band magnitude then decreases by the amount $\Delta V \sim 0.35$, 
returning almost to its initial brightness before JD 2453540. 
This behavior defines a $\sim 630$ day wide maximum delimited by 
two minima at JD 2453540 and JD 2454172.
Since then, $\eta$~Car holds a constant brightness (neglecting
shallower fluctuations), and 
toward the end of the 2008 observing season, appears to be starting 
to brighten slightly.

\begin{figure*}
\centering
        {\includegraphics[width=18.0cm,angle=0]{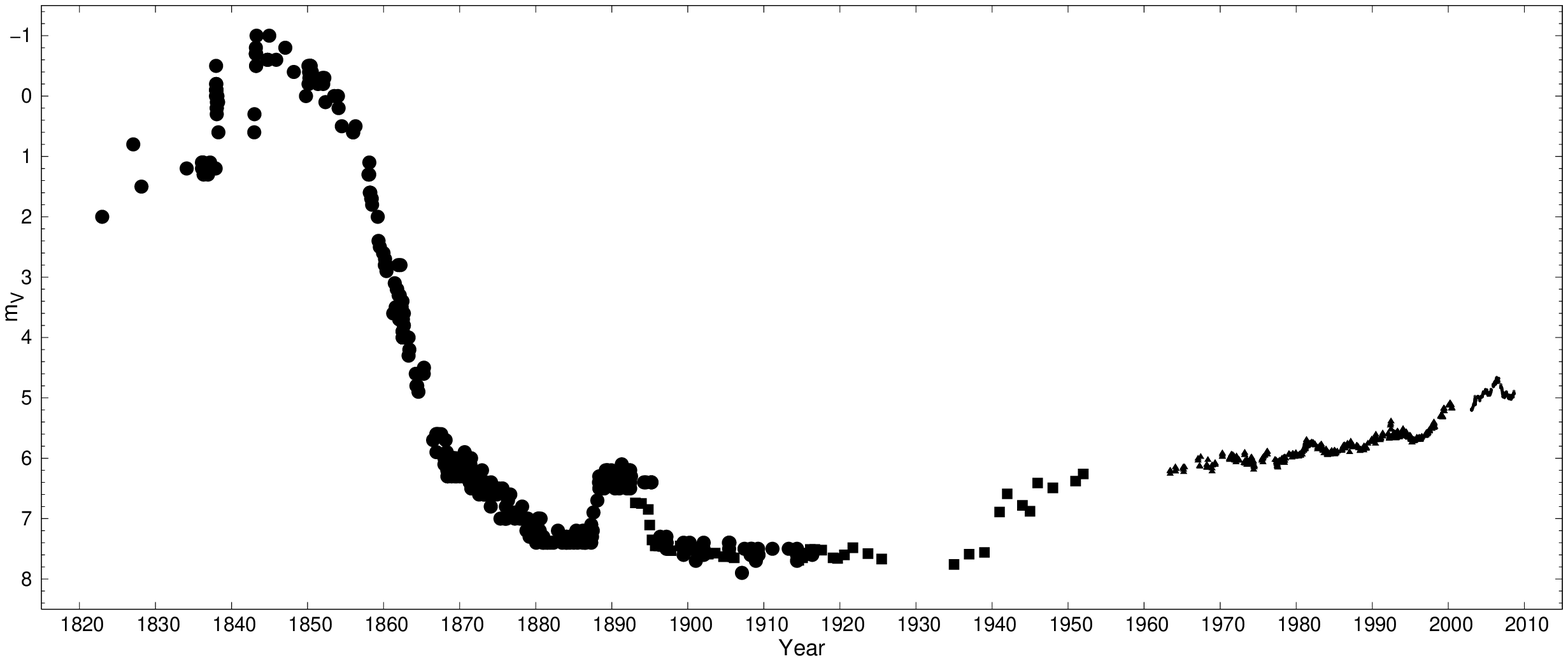}}
        \caption{
                Light curve of $\eta$ Car since 1820 until 2008 including
                visual ({\Large $\bullet$}), photographic
                ($\blacksquare$), photoelectric ($\blacktriangle$) and our $V$ CCD
                ({\tiny $\bullet$}) observations.
                The maximum peak reached in 2006 and the consecutive decline
                are clearly visible. 
                }
        \label{historic}
\end{figure*}

Figure~\ref{plots} illustrates that fluctuations occur for a significant part of the orbital
cycle. 
Most of them are due to the S Dor nature of $\eta$~Car, although
some others must be related to its binarity.
Large scale features, such as secular brightening or fading, become more evident
in a global context. 
To illustrate this, we compiled the light curve of $\eta$~Car from 1820
to 2008, which is depicted in Fig.~\ref{historic}. 
We have plotted all the historical visual estimates obtained between 1822 and 1916, compiled by \citet{Frew}. 
We decided to extend the curve through to the present exclusively
with instrumental measurements, i.e. photographic, photoelectric, and CCD. 
They included our $V$ observations (see Fig.~\ref{plots}) and the data published by 
\citet{Hoffleit,O'Connell,Feinstein67,Feinstein74,
Sterken96,Sterken99,vG06}.
The photographic data \citep{Hoffleit,O'Connell} 
were transformed to visual magnitudes using a zeropoint of -0.85 mag, to match those of \citet{Frew}.
While the major part of the photoelectric magnitudes were extracted 
with apertures similar to the one adopted in this paper, the data from
\citet{Sterken96,Sterken99} were
derived for a smaller aperture.
Therefore, we applied zeropoints to these two datasets of $-0.15$
and $-0.20$ mag, respectively. 
The time series, as used to compile Fig.~\ref{historic}, are given in
Table 3\footnote{Table 3 is available only in electronic form at the CDS.}.

As is evident in Fig.~\ref{historic}, $\eta$~Car has been increasing  in brightness since about 1940. 
It changed the increasing rate around 1950, and more recently in 1997,
reaching the maximum $V \sim 4.7$ in 2006\footnote {It is worth mentioning
that the last time $\eta$~Car showed that $V$ magnitude was in early 1860s
in the fading phase after the {\em``Great Eruption''}.}.
 The object later faded and then remained at a brightness of  $V \sim 5$, showing no secular
variations.

\section{Concluding remarks}
In the following, we outline the main results derived from our photometry since
January 2003:
\begin{itemize}

\item The system maintained its tendency to brighten in the four photometric bands until 2006.35, when the maximum brightness in 150 years was
    reached.
    
\item After this maximum in 2006.35, the brightness of $\eta$~Car started to decline.
Since 2007.2, the $BVRI$ light curves have remained at almost constant brightness, showing a slight brightness rise in the days of writing of this report.
As we could expect considering the events of 2003, such an increase is
probably associated with an upcoming periastron passage.

\item Just after the last periastron passage in 2003.5, the photometry
  displayed a wide maximum, called an ``egress-maximum'' by \citet{vG06}, which finished at a clearly evident minimum.

\item The maximum peak was reached shortly after the orbital phase $\phi
  \sim 0.5$. This encourages us to relate it with some type of phenomena
  occurring during apastron passage.
The light fluctuation detected about that maximum peak is quite symmetrical, 
covering 30\% of the period. 
This would be consistent with the flux variations that occur after the
periastron passage, reported by \citet{Duncan03,Damineli08b}.
Nevertheless, we cannot discount a possible coincidence that an S Dor phase just happened to occur at that particular orbital moment.

\item The time at which a minimum occurred was estimated from the 2003.5 event, and by adopting a 2022.7 days period, the next optical photometric dip might be
expected in January 28, 2009. 
The onset of the 2009-event might have occurred about one or two months before.

\end{itemize}

We have provided dense and well defined multiband
light curves of $\eta$~Car + Homunculus, covering the entire orbital
period of the proposed binary system. 
This consistent and homogeneous data set provides invaluable information for
further modeling, analysis, and interpretation, which we expect to enhance our understanding of the physical nature of $\eta$ Car. 

\section{Present and future prospects}
Our optical monitoring of $\eta$~Car remains in process.
Our up-to-date light curves can be found at our web-page
\footnote {{\it http://etacar.fcaglp.unlp.edu.ar/}}.
The following improvements are being performed:

\begin{itemize}

\item Different aperture sizes and point spread function fittings will be tested in extracting the
instrumental magnitudes and analyzing the contribution to the light curves of
different areas of the Homunculus. 

\item $H\alpha$ and Johnson $U$ filters have been included in our observation
program at La Plata and CASLEO\footnote {Complejo Astron\'omico el
Leoncito, San Juan, Argentina} observatories, respectively.

\item We propose to complement our daily monitoring with all-night long
  observations of $\eta$~Car, certainly from Dec 2008, to
  ensure that we will be able to derive the light curves of the anticipated 2009 ``eclipse-like'' event in the $U, B, V,
  R, I$, and $H\alpha$ bands. 

\end{itemize}

\begin{acknowledgements}
The authors acknowledge the authorities of the FCAG-UNLP
for the use of the observational facilities at La Plata Observatory.
We acknowledge the participation of the following astronomers and
students of the FCAG-UNLP during the observations:
Violeta Bazzano,
Ayeray Bonansea,
Dr. Roberto Gamen,
Maximiliano Haucke,
Matilde Iannuzzi,
Lorena Mercanti,
\'Angeles Molin\'e,
Alejo Molina Lera,
Cintia Peri,
Cecilia Scalia,
Lautaro Simontacchi,
and
M. Florencia Teppa Pania.
Their presence during observations was of a significant help and they are gratefully
acknowledged. 
We warmly thank Federico Bareilles for several, important contributions on the computational resources and eng. Ezequiel
Garcia for the maintenance of and improvements to the telescope and its
equipment. 
We thank the referee, Dr. C. Sterken, for valuable suggestions, which
improved  the presentation of this paper.
We remember Reinhardt Glinschert, our telescope mechanician,
who passed away last January 2008, and our ``maestra'' Virpi Niemela, who
inspired this program and left us in December 2006.

\end{acknowledgements}

\bibliographystyle{aa}
\bibliography{aav5}

\end{document}